\documentclass[manuscript,nonacm]{acmart} %anonymous,
\RequirePackage{hyperref}
\RequirePackage{hyperxmp}
\usepackage{etex}
\usepackage{soul}
\usepackage{graphicx,comment}
\usepackage{amsmath}
\usepackage{algorithm}
\usepackage{verbatim}
\usepackage{multirow}
\usepackage{booktabs}
\usepackage{makecell}
\usepackage{longtable}
\usepackage{enumitem}
\usepackage{xspace}
\usepackage{bm}
\usepackage{longtable}
\usepackage{caption}
\usepackage{adjustbox}
\usepackage{tcolorbox}

\usepackage{color, colortbl}
\definecolor{codegreen}{rgb}{0,0.6,0}
\definecolor{codegray}{rgb}{0.5,0.5,0.5}
\definecolor{codepurple}{rgb}{0.58,0,0.82}
\definecolor{backcolour}{rgb}{0.95,0.95,0.92}
\definecolor{LightCyan}{rgb}{0.88,1,1}
\definecolor{LightRed}{RGB}{255, 204, 203}
\usepackage{amsmath}

\makeatletter
\def\adl@drawiv#1#2#3{%
        \hskip.5\tabcolsep
        \xleaders#3{#2.5\@tempdimb #1{1}#2.5\@tempdimb}%
                #2\z@ plus1fil minus1fil\relax
        \hskip.5\tabcolsep}
\newcommand{\cdashlinelr}[1]{%
  \noalign{\vskip\aboverulesep
           \global\let\@dashdrawstore\adl@draw
           \global\let\adl@draw\adl@drawiv}
  \cdashline{#1}
  \noalign{\global\let\adl@draw\@dashdrawstore
           \vskip\belowrulesep}}
\makeatother

\usepackage{pifont}% http://ctan.org/pkg/pifont

\usepackage{minibox}
\newcounter{observcntr}

\newcommand{\subheading}[1]{\noindent{\textbf{#1}}}

\usepackage{array}
\newcolumntype{L}[1]{>{\raggedright\arraybackslash}p{#1}}

\AtBeginDocument{%
  }

\begin{document}

%\title{Rethinking Cybersecurity: Red and Blue Teaming in the LLM Era}
%\title{New Ways to Consider Agentic AI Security: A Human Society-Inspired 4C Framework}
\title{Human Society-Inspired Approaches to Agentic AI Security: The 4C Framework}

% \author[inst1]{Alsharif Abuadbba}
% \author[inst1]{Nazatul Sultan}
% \author[inst1]{Surya Nepal}

% \affiliation[inst1]{%
%   \institution{CSIRO's Data61}
%   \city{Sydney}
%   \state{NSW}
%   \country{Australia}
% }

% \author[inst2]{Sanjay Jha}
% \affiliation[inst2]{%
%   \institution{University of New South Wales, Sydney}
%   \city{Sydney}
%   \state{NSW}
%   \country{Australia}
% }

\author{Alsharif Abuadbba}
\affiliation{%
  \institution{CSIRO's Data61}
  \country{Australia}}
\email{sharif.abuadbba@data61.csiro.au}

\author{Nazatul Sultan}
\affiliation{%
  \institution{CSIRO's Data61}
  \country{Australia}}
\email{nazatul.sultan@data61.csiro.au}

\author{Surya Nepal}
\affiliation{%
  \institution{CSIRO's Data61}
  \country{Australia}}
\email{surya.nepal@data61.csiro.au}

\author{Sanjay Jha}
\affiliation{%
  \institution{University of New South Wales, Sydney}
  \country{Australia}}
\email{sanjay.jha@unsw.edu.au}

\renewcommand{\shortauthors}{Alsharif Abuadbba et al.}

\maketitle
AI is moving from domain-specific autonomy in closed, predictable settings to large-language-model-driven agents that plan and act in open, cross-organizational environments. As a result, the cybersecurity risk landscape is changing in fundamental ways. Agentic AI systems can plan, act, collaborate, and persist over time, functioning as participants in complex socio technical ecosystems rather than as isolated software components. Although recent work has strengthened defenses against model and pipeline level vulnerabilities such as prompt injection, data poisoning, and tool misuse, these system centric approaches may fail to capture risks that arise from autonomy, interaction, and emergent behavior. This article introduces the 4C Framework for multi-agent AI security, inspired by societal governance. It organizes agentic risks across four interdependent dimensions: Core (system, infrastructure, and environmental integrity), Connection (communication, coordination, and trust), Cognition (belief, goal, and reasoning integrity), and Compliance (ethical, legal, and institutional governance).
By shifting AI security from a narrow focus on system centric protection to the broader preservation of behavioral integrity and intent, the framework complements existing AI security strategies and offers a principled foundation for building agentic AI systems that are trustworthy, governable, and aligned with human values.

\section{Introduction: Why Agentic AI Security Needs a New Lens}
Artificial intelligence is entering a new phase. Recently, Large Language Models (LLMs) have demonstrated strong capabilities in question answering, summarization, code generation, and multi step reasoning across many domains. Despite significant advances, AI deployments for many years have remained largely confined to predictive use cases, such as credit-risk scoring, fraud detection, and recommendation ranking. In most real‑world deployments, modern LLMs, ChatGPT among them, remain fundamentally reactive: they receive a prompt, generate a response, and then fall silent, ceding the subsequent action to the user or to the software scaffolding that holds them. A newer direction, often called agentic AI, is beginning to shift systems from reaction to action. Agentic systems can pursue goals by planning steps, calling tools and APIs (for example, searching internal databases, reading documents, querying logs, or running code), using data, retaining state across interactions (memory), and coordinating with humans or other agents with less step by step supervision. Vendors (e.g., Amazon Bedrock Agents) and open-source frameworks (e.g., LangGraph) are already productizing agentic workflows. %Major vendors are already productizing this direction, for example through Amazon Bedrock Agents and AgentCore, Google Agentspace, Microsoft Agent Framework, and Salesforce Agentforce. In parallel, open source ecosystems provide lightweight frameworks such as AutoGPT, CrewAI, and LangGraph that make agentic and multi-agent workflows broadly programmable.

As agentic systems mature, they start to look less like tools and more like participants in a socio‑technical society, a landscape where human intent and technological capability intermingle to shape what happens next. They interact, coordinate, and shape one another’s behavior, making autonomous decisions whose effects ripple from digital systems into real operations, cross-organizational settings, and real-world consequences. This evolution fundamentally changes the nature of security. To date, much of the work on securing LLMs and agentic AI has remained predominantly system‑centric, concerned with safeguarding the model itself and the technical substrate that underpins it, from software layers to the infrastructure on which it runs. This includes guarding against  prompt injection, data poisoning, tool misuse, sandbox escapes, and model extraction attacks. Influential frameworks and benchmarks such as InjecAgent, MINJA, and the OWASP GenAI Top 10 have been instrumental in codifying these threats and fortifying the technical pipeline around agents, from memory modules and tool‑calling interfaces to APIs, retrieval chains, and the runtimes that orchestrate them. Yet these approaches largely cast agents as software artifacts rather than as autonomous social actors. In doing so, they risk overlooking forms of harm that emerge through interaction and behavior such as social engineering. For instance, GPT‑4 has demonstrated the ability to impersonate a vision impaired user to persuade a human to solve a CAPTCHA on its behalf, bypassing the reCAPTCHA not through technical exploitation, but through deception. This points to a growing class of risks shaped not by isolated system defects but by strategic interaction, autonomous behavior, and an agent’s capacity to influence others.

%As agentic systems evolve, they increasingly resemble participants in a socio-technical society—interacting, coordinate, and influence one another, and make autonomous decisions that propagate from digital environments into the physical world with real-world consequences. This evolution fundamentally changes the nature of security. To date, most research on LLM and agentic AI security has remained \textit{system-centric}, focusing on protecting the model and its execution environment. This includes preventing prompt injection, data poisoning, tool misuse, sandbox escapes, and model extraction attacks. Influential frameworks and benchmarks—such as InjecAgent, MINJA, and the OWASP GenAI Top 10—have played a critical role in formalizing these threats and hardening the technical pipeline around agents, including memory modules, tool call mechanisms, APIs, retrieval chains, and agent runtimes. Although these approaches provide rigorous protection at the component and infrastructure levels, they largely treat agents as \textit{software artifacts rather than autonomous social actors}. Their primary objective is to ensure the robustness of the execution layer, not to characterize or constrain emergent behavior arising from interaction, coordination, or strategic adaptation. As a result, a growing class of risks remains underexplored—risks that do not originate solely from isolated system defects, but also from behavior, interactions, and autonomy.

In this paper, we argue that agentic AI introduces forms of threat that echo those found in human societies: patterns of behavior implicitly absorbed during training and expressed through autonomous interaction. Agents can form brittle or drifting beliefs, influence or manipulate their peers, negotiate or collude toward unintended goals, or strategically evade constraints to maximize outcomes. Such behaviors fall largely outside existing vulnerability taxonomies, yet they can ripple across multi‑agent ecosystems, producing cascading failures and systemic misalignment rather than isolated faults. Human societies have long confronted analogous challenges using layered defenses such as biological immunity, cognitive safeguards, social norms, ethical principles, and legal governance. As agentic AI systems expand from individual models to interconnected populations of agents, we contend that they will require safeguards of a similarly layered character. To that end, we introduce the 4C Framework for Multi‑Agent AI Security, broadening the scope of security beyond technical robustness to encompass four complementary dimensions.
(1) \underline{\textbf{Core}} – the integrity of the agent's digital body, including the infrastructure that runs it and the environment it operates in. 
(2) \underline{\textbf{Connection}} – how agents communicate, coordinate and influence one another.
(3) \underline{\textbf{Cognition}} – how beliefs, goals, and plans are formed and updated.
(4) \underline{\textbf{Compliance}} – how agent behavior stays within ethical, legal, and institutional boundaries. Together, these layers shift security beyond traditional vulnerabilities toward agency, interaction, and collective behavior. The 4C Framework complements system-level work by highlighting risks that arise when AI becomes not just a model, but a population of interacting agents embedded in human–machine ecosystems.
% \begin{itemize}
% \item Core – the integrity of the agent's digital body, including the infrastructure that runs it and the environment it operates in 
% %the foundational integrity of the agent’s digital ``body'', the infrastructure that hosts / supports it and its operation in the cyber-physical-social environment
% \item Connection – how agents communicate, coordinate and influence one another % how agents communicate, cooperate and influence
% \item Cognition – how beliefs, goals, and plans are formed and updated %how goals, reasoning, and internal beliefs are formed
% \item Compliance – how agent behavior stays within ethical, legal, and institutional boundaries %how agents stay aligned with the external ethical, legal, and institutional boundaries
% \end{itemize}

%Together, these layers shift security thinking beyond traditional vulnerabilities toward a deeper understanding of agency, interaction, and collective behavior. The 4C Framework complements existing system level work by illuminating the risks that arise when AI becomes not just a model but a population of interacting agents embedded within human machine ecosystems.

%Together, these layers shift security thinking beyond traditional vulnerabilities toward a deeper understanding of agency, interaction, and collective behaviour. The 4C Framework complements existing system-level work by illuminating the emerging risks that arise when AI becomes not just a model—but a community of interacting agents within complex human–machine ecosystems.
\vspace{-.3cm}
\begin{tcolorbox}[
  title={Key Insights},
  colback=white,
  colframe=black,
  fonttitle=\bfseries,
  boxrule=0.8pt,
  left=6pt,
  right=6pt,
  top=6pt,
  bottom=6pt
]
\begin{enumerate}
  \item Agentic AI reframes cybersecurity risk by shifting the focus from isolated technical exploits to failures that emerge from behavior and interaction, especially in cross-organizational multi-agent systems where small errors can propagate and cascade.
  \item Securing models, tools, and execution environments is necessary but not sufficient, because many agentic risks stem from belief formation, influence, delegation, and long-horizon autonomy rather than from traditional vulnerabilities.
  \item We offer a societal governance perspective on agentic AI security, arguing that genuine trustworthiness cannot be achieved through technical controls alone. We develop this view in the 4C Framework, which brings together Core, Connection, Cognition, and Compliance as interdependent layers of governance.
\end{enumerate}
\end{tcolorbox}

\section{The Evolution of Intelligent Automation to Multi-Agent AI}
\begin{figure}[!t]
    \centering
    \includegraphics[width=.9\linewidth]{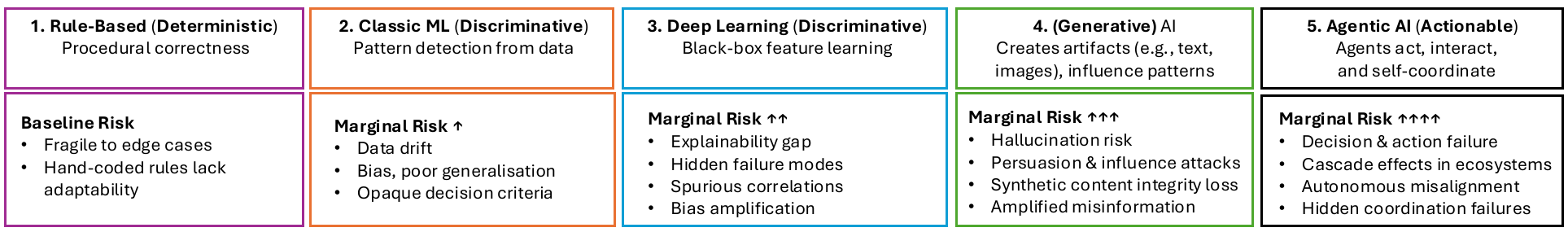}
    \caption{Marginal risk across the evolution from automation to multi-agent AI.}
    \Description{Marginal risk across the evolution from automation to multi-agent AI.}
    \label{fig:marginalrisk}
    %\vspace{-.5cm}
\end{figure}

Over three decades, digital automation has evolved from scripted workflows to statistical learning, then to deep learning, to generative models, and ultimately to autonomous multi‑agent systems. Each stage enlarged capability, but it also widened the attack surface and deepened system dependencies. Figure \ref{fig:marginalrisk} shows how the marginal risk, the additional risk incurred when moving from one stage of automation to the next, can rise sharply as systems move from passive prediction to autonomous action and coordination between agents.
For example, an incident response workflow that once only flagged anomalies may now open tickets, query logs, and trigger remediation steps through tools, which increases the consequences of mistakes and misuse.

%Over three decades, digital automation has advanced from scripted workflows to statistical learning, deep learning, generative models, and now autonomous multi-agent intelligence. Each stage expanded capability—yet also expanded the attack surface and systemic dependencies. Figure~\ref{fig:marginal_risk} illustrates how marginal risk increases nonlinearly as systems evolve from passive prediction to autonomous action and inter-agent coordination.

\subsection{From Automation to Multi-Agent Intelligence}

The trajectory of digital automation can be understood across five phases:

\subheading{1. Rule-Based Automation.}  
Deterministic systems execute predefined workflows using hand-crafted rules. Risks are primarily operational, including brittleness, misconfiguration, and poor handling of edge cases.
%\cite{vander2018}
\subheading{2. Classic Machine Learning.}  
Supervised and unsupervised machine learning introduced statistical prediction for tasks such as classification, regression, detection, and ranking. These methods were largely discriminative, learning decision boundaries or scoring functions from human \emph{annotated} data and engineered features. Models were typically trained offline and remained static after deployment, making them sensitive to distributional drift. While risk increased due to bias and dataset shift, failures were usually localized, interpretable, and confined to specific domains.
%\cite{bishop2006}
%These methods were largely discriminative, learning decision boundaries or scoring functions from \emph{human-annotated} data and manually engineered features \cite{bishop2006pattern}. Models were typically trained offline and remained static after deployment, making them sensitive to distributional drift. While marginal risk rose due to bias and dataset shift, failures typically remained localized, interpretable, and confined to their specific domains.

\subheading{3. Deep Learning.}  
%The term ``deep'' in deep learning refers to the use of multi-layer neural network architectures. Like classic machine learning, deep learning models remained predominantly discriminative, optimized to map inputs to labels, scores, or probabilities by separating classes in high-dimensional representation spaces. What changed was how features were obtained: deep neural networks reduced reliance on manual feature engineering by learning hierarchical representations directly from data through successive layers of abstraction \cite{lecun2015deep, goodfellow2016deep}. This shift enabled major breakthroughs in computer vision, speech recognition, and sequence modelling. However, as representations became embedded within increasingly deep architectures, marginal risks arising from opaque decision criteria increased, making models harder to audit and exposing hidden failure modes. Despite their expressive power, these systems remained fundamentally passive: they could predict, but they could neither plan nor act.
Deep learning uses multilayer neural networks that learn meaningful patterns directly from data, reducing the need for manually designed input cues~\cite{lecun2015deep}, \cite{goodfellow2016deep}. This transition unlocked major progress in vision, speech, and language, but it also brought decision processes that were opaque and failure modes that were hard to audit. However, these systems remained essentially passive. They could predict, but they could not plan or take actions.

%The term “deep” in deep learning refers to the use of multi-layer neural network architectures that learn hierarchical representations directly from data, reducing reliance on human feature annotation in earlier machine learning systems \cite{lecun2015deep, goodfellow2016deep}. Like classic machine learning, these models remained predominantly discriminative, mapping inputs to labels or scores in high-dimensional representation spaces. This shift enabled major advances in computer vision, speech recognition, and sequence modeling, but also increased marginal risk by introducing opaque decision criteria and hidden failure modes that are difficult to audit. Despite their expressive power, deep learning systems remained fundamentally passive: they could predict, but neither plan nor act.

\subheading{4. Generative AI and LLM Co-Pilots.} 
% Between 2021 and 2023, LLMs introduced reasoning-like behaviour, content generation, and in-context learning \cite{openaiGPT42024}. These systems marked a shift from predominantly discriminative models to generative models, which learn the underlying distribution of language and can produce novel text, code, and structured outputs conditioned on context, rather than merely assigning labels or scores. Architecturally, LLMs leveraged transformer-based designs that emerged in 2017, while diffusion models emerged in parallel as the dominant generative paradigm for images and, later, video.
% A defining characteristic of these models is their domain-agnostic nature: a single system can operate across diverse domains without task-specific retraining. While this generality enabled rapid deployment and broad productivity gains, it also introduced new marginal risks, as errors, hallucinations, or persuasive misuse could propagate across multiple high-stakes domains. Despite these advances, the models largely remained “in the box”: humans provided prompts, models generated responses, and humans interpreted and executed downstream actions. Risks increased—including hallucinations, persuasive misuse, and synthetic-content integrity challenges—but remained partially moderated by human-in-the-loop workflows and the absence of autonomous action.
In the early 2020s, large language models made it easy for computers to generate meaningful text and code from a prompt, often displaying reasoning like behavior. This marked a shift from earlier AI systems that mainly labeled or scored inputs to ones that could produce new content shaped by context. Many of these new models relied on transformer based designs, and related generative methods soon enabled the creation of images and video as well. One model can support many tasks without retraining, which helped adoption spread quickly. At the same time, new risks emerged, including hallucinations, errors in high stakes settings, and persuasive misuse. Even so, most generative systems stayed within the software interface, with human-in-the-loop deciding the next step and retaining final control. In the incident response example, the model might draft a summary or suggest a query, but a person still runs the commands and applies the fix.

%In the early 2020s, LLMs introduced reasoning-like behavior, content generation, and in-context learning, marking a shift from predominantly discriminative models to generative systems \cite{openaiGPT42024}. Rather than assigning labels or scores, these models learn the distribution of language and produce novel text, code, and structured outputs conditioned on context. Architecturally, LLMs build on transformer-based designs introduced in 2017, with diffusion models emerging in parallel as the dominant generative approach for images and later video. A defining feature of these models is their domain-agnostic generality: a single system can operate across diverse tasks without retraining. While this enabled rapid deployment and broad productivity gains, it also introduced new marginal risks, as hallucinations, errors, and persuasive misuse could propagate across high-stake domains. Despite these risks, generative models remained largely "“in the box,” with humans providing instructions and performing downstream actions, moderating impact through human-in-the-loop control.

\subheading{5. Agentic and Multi-Agent AI.}  
AI is shifting from stand-alone generative models to agentic systems. Where earlier deep learning reduced the need for manual feature engineering, agentic AI reduces the need for humans to manually execute routine workflows. An agentic system takes a natural-language request, decomposes it into steps, uses an LLM to plan, invokes external tools (APIs, databases, ticketing systems), and carries out actions with logging and guardrails. For example, a service-desk assistant can turn ``I can't access my account" into a short plan: verify identity, check recent authentication failures, confirm account status, trigger an approved reset or restoration runbook, and record the actions taken in the ticket. Unlike earlier rule based agents, modern agentic AI is adaptive and can reason, plan, and act in open environments using general-purpose foundation models (large models trained on broad data)~\cite{wang2025large}, \cite{abuadbba2025promise}. As these systems grow from a single agent to many, new risks appear. In multi-agent AI ecosystems, agents negotiate, share state such as task context and intermediate results, and pursue objectives in parallel. This enables scalable automation but also increases systemic risk: cascading failures, misaligned actions, and emergent group behavior can arise from agent interactions.

\begin{figure}[!t]
  \centering
  {\includegraphics[width=0.6\linewidth]{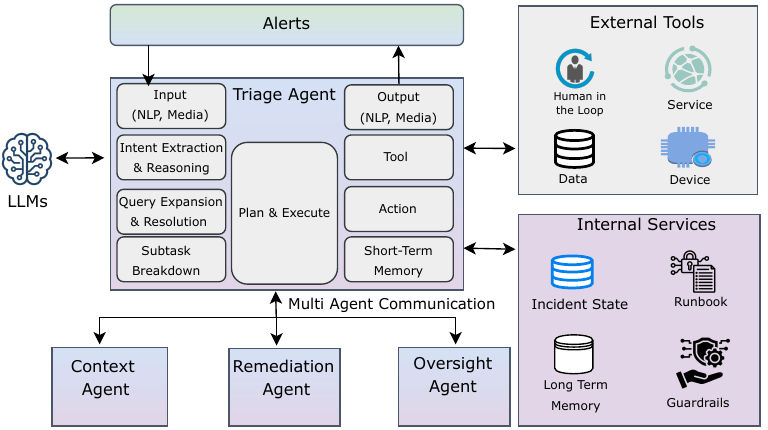}}
  \caption{Example Architecture: Multi-Agent Agentic AI System}
  \Description{Example Architecture: Multi-Agent Agentic AI System}
  \label{fig:multiagent}
  %\vspace{-.6cm}
\end{figure}

\subsection{Modern Multi-Agent AI: Definition and Security Challenges}
Multi-agent AI systems consist of multiple autonomous agents, each capable of perception, reasoning, memory, and action, that work together and interact with their environment across organizational boundaries to pursue individual or shared goals. Figure \ref{fig:multiagent} presents a reference architecture example for such systems. Because agents coordinate through communication and task handoffs, system-level behaviors can emerge that are not reducible to any single component, including cooperation, competition, manipulation, or other unexpected dynamics. To make this concrete, consider a multi-agent incident response setup. A triage agent monitors alerts and opens an incident record, a context agent pulls logs and asset context, a remediation agent proposes (or executes) an approved runbook, and an oversight agent enforces approvals before any high-impact step. These agents coordinate through message handoffs and shared incident state (for example, evidence bundles, recommended actions, and approval decisions), so security depends on the integrity of communication, delegation, and shared context, not just on any single model. In this sense, multi-agent deployments begin to resemble distributed socio-technical organizations rather than isolated AI models.
\par 
As agentic workflows span multiple agents, shared data flows, tools, and across organizational boundaries, security approaches centered on protecting a single model or execution pipeline become inadequate \cite{zhang2025landscape}. Many risks arise not only from isolated failures but also from interaction effects, including cascading errors, persuasion or manipulation, malicious or mistaken task handoffs, and unsafe tool use amplified by long-horizon planning. In such systems, risk becomes a property of the ecosystem itself, shaped by how agents coordinate, influence one another, and act over time. \textit{This shift exposes a fundamental gap: existing security frameworks do not fully capture how threats emerge, propagate, or compound in multi-agent systems. %Addressing these risks requires security models that account for agent intent, interaction, and governance, motivating the 4C Framework introduced next.}
Addressing these risks requires security models that account for agent intent, interaction, and governance, motivating the 4C Framework.}

\begin{figure}[!h]
    \centering
    \includegraphics[width=0.6\linewidth]{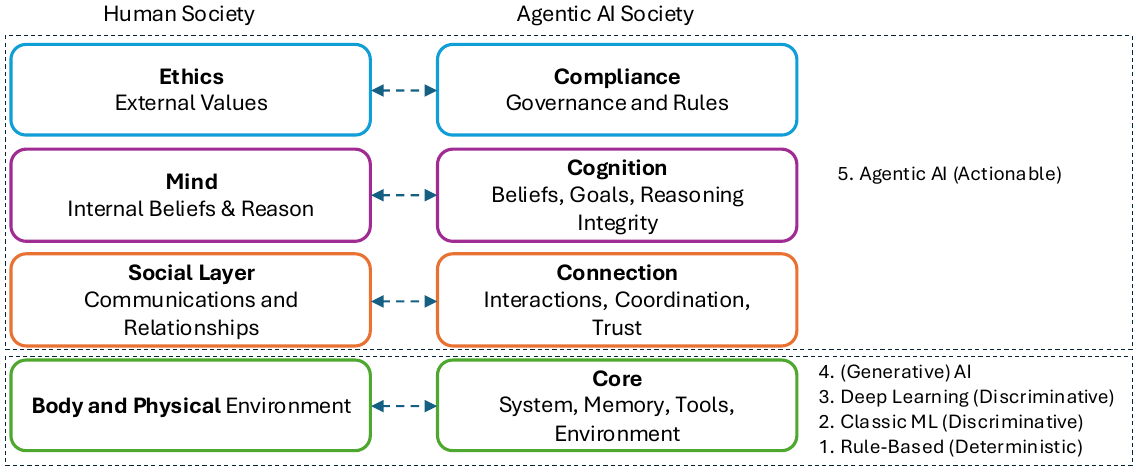}
	\caption{Human Society Analogy → Agentic AI Society 4C Mapping}		
    \label{fig:4c-system}
    \vspace{-.3cm}
 \end{figure} 

\section{The 4C Framework for Multi-Agent AI Security: A Human-Inspired Perspective}
\begin{figure}[!t]
    \centering
    \includegraphics[width=0.8\linewidth]{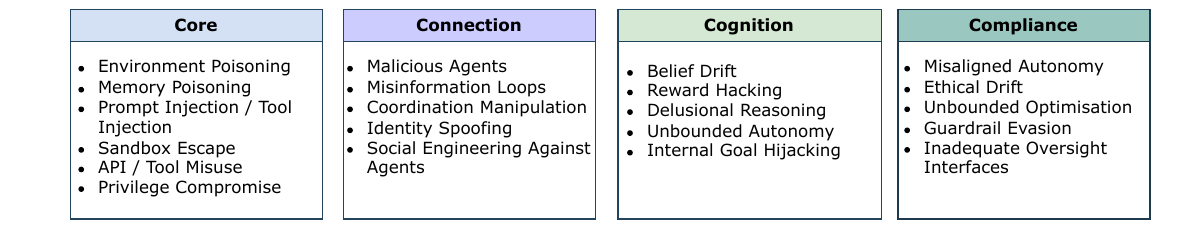}
	\caption{Illustrative examples of threat categories across the 4C Framework}		
    \label{fig:security-risks}
    %\vspace{-.5cm}
 \end{figure}

Securing agentic AI requires acknowledging that these systems are not single models. They are assembled from multiple components that sense, reason, act, and coordinate across real environments, often over extended workflows. As a result, their risks span not only model and infrastructure failures but also interaction failures and governance failures. To organize this landscape, we propose the \textbf{4C Framework for Multi-Agent AI Security} depicted in Figure~\ref{fig:4c-system}, a human-inspired lens with four connected layers: Core, Connection, Cognition, and Compliance. Each layer draws an analogy to human systems, from the body and mind to social interaction and governance, and highlights the threats and mitigations most relevant at that level. Figure \ref{fig:security-risks} summarizes the representative threat categories in the four layers. We use the multi-agent incident response example introduced earlier to illustrate the framework throughout the paper.
%Securing Agentic AI requires understanding that these systems are not monolithic models, but distributed entities that perceive, reason, act, and interact across complex environments. Their vulnerabilities are layered similarly, extending from inherited component risks to emergent social and governance challenges. To reason about this broader landscape, we propose the \textbf{4C Framework for Multi-Agent AI Security}---a human-inspired model mapping four interdependent dimensions of safety and control: \textit{Core, Connection, Cognition, and Compliance}. Each dimension draws an analogy to human systems, from the body and mind to social interaction and governance, providing an integrative view of intelligent agency, and the threats and mitigations most salient at each layer. Figure \ref{fig:security_risks} summarizes the representative threat categories in the four layers. We use the multi-agent incident response setup example, described earlier, to illustrate all four layers in rest of the paper. 

\subsection{Core: System Component and Environmental Security}

\subheading{Definition and Scope.} The Core layer forms the system foundation that allows an agent to exist and operate. It includes the runtime and execution environment where the agent runs, the memory stores that hold what it can retain and reuse, the tool and API interfaces it can call, the data pipelines it relies on, and the permissions and shared resources that determine what it can access. This is the layer that turns an LLM from ``generate text" into ``take actions": retrieve context, call tools, store intermediate state, and produce outputs that can trigger real changes in other systems. In human terms, Core is the agent's ``digital body": the underlying machinery and operating conditions that enable action, and along with  basic safeguards that prevent takeover. In the incident response multi-agent example, the Core layer is what allows the triage agent to open an incident record, the context agent to pull logs and asset details, the remediation agent to run an approved step in a runbook, and the oversight agent to query the ticketing system to verify approvals. If the Core is weak, reasoning quality does not matter: poisoned data, compromised tools, or overly broad permissions can drive unsafe actions.
The \textit{scope} of the Core layer is deliberately narrow. It governs how an agent runs and what it can access, not why it acts, what it believes, or whether its behavior is socially or ethically acceptable. Those questions belong to the other layers, which we describe next.

%The Core layer represents the foundational system level infrastructure that enables an agent to exist and operate. In agentic environments, this infrastructure allows LLMs to move beyond single-shot text generation to goal-directed action. It encompasses the runtime, execution environment, memory stores, tool interfaces, APIs, data pipelines, permissions, and shared resources that collectively support agent behavior. Through the interaction of LLMs, tools, and memory, a user request can be transformed into concrete actions such as planning, retrieving context, invoking tools, and producing output. For example, Microsoft Security Copilot\footnote{\url{https://learn.microsoft.com/en-us/azure/sentinel/sentinel-security-copilot}} integrates LLM reasoning with governed organizational security data from platforms such as Microsoft Sentinel and Defender, allowing workflows such as summarizing incident context or generating search queries. Such systems demonstrate how controlled tool access and data integration enable agent-like capabilities that extend well beyond text-only assistance. At this level, the Core layer is responsible for the mechanical correctness and integrity of execution. Failures here compromise the agent holistically, rendering higher-level safeguards ineffective. The scope of the Core layer is intentionally narrow: it addresses how an agent runs, not why it acts, what it believes, or whether its behavior is socially or normatively appropriate. These concerns are explicitly deferred to higher layers of the framework.

\subheading{Threat Categories and Examples.}
Threats in the Core layer are largely technical and environmental, and they are well documented in existing AI security research (e.g., \cite{kong2025,Deng2025}). These threats  arise from vulnerabilities in the agent’s execution substrate and supporting infrastructure, rather than from higher-level reasoning or interaction. Threats are amplified by the expanded attack surface introduced by tool use, memory persistence, and execution environments. We briefly discuss a few critical threats in the Core layer. (1) \textit{Environmental poisoning} occurs when the agent is induced to ingest malicious or misleading information from external entities, such as compromised tools, databases, or retrieved web content \cite{Das2025}. In our incident-response example, this could occur if a context agent pulls logs from a compromised source that systematically conceals signs of intrusion, causing the system to under react or to choose the wrong remediation step. (2) \textit{Memory poisoning} occurs when an attacker inserts false or compromised information into the agent's memory to bias subsequent behavior \cite{zou2025poisonedrag}. In incident response, an injected memory entry such as ``host A is already isolated" or ``this alert type is always benign"  can lead to repeated suppression of valid escalations, even when the underlying tools and models remain unchanged. (3) \textit{Other threats}, including prompt or tool injection, where an attacker manipulates the agent’s instructions, sandbox escape, where execution barriers fail, misuse of APIs or tools, where legitimate capabilities are abused, and privilege compromise, where access rights are escalated, can further erode the reliability, integrity, and safety of agentic functions \cite{narajala2025}. %These threat categories constitute the primary focus of established security frameworks such as InjecAgent, MINJA, and the OWASP GenAI Top 10. Each offers useful taxonomies and mitigation strategies to harden agent runtimes and execution environments.

\subheading{Key Remarks and Mitigation Implications.}
Core security is fundamentally  a matter of digital physiology and hygiene. It depends on clearly specified components, strong guarantees of integrity, and a tightly controlled operating environment. In practice, this means sandboxing agents and tools, enforcing least-privilege access, validating tool inputs and outputs rather than trusting them by default, and monitoring for patterns indicative of environmental poisoning, memory manipulation, or boundary violations. For an incident‑response agent, this translates into disciplined boundaries: allowlisting only the tools it may invoke, granting credentials scoped to the smallest set of necessary actions, accepting critical evidence only when it arrives signed or carrying verified provenance, and maintaining a strict divide between tools that merely \textit{read} (log queries) and those that can \textit{act} (host isolation, credential rotation). Although securing the Core cannot eliminate every risk, it provides a stable, trustworthy substrate on which safer mechanisms for connection, cognition, and compliance can be built.  As argued in \cite{potter2025}, Core-layer security improves when agents are built with secure-by-design and secure-by-construction practices applied across the system lifecycle.%Concretely, this includes threat modeling and secure SDLC controls, hardened sandboxed execution, supply-chain integrity checks for tools and dependencies, and continuous monitoring with incident-response playbooks for abnormal agent behavior.

\subsection{Connection: Communication and Social Security}

\subheading{Definition and Scope.} In the Core layer, we considered what an agent depends on to run such as its models, memory, tools, and execution environment. The Connection layer shifts the focus from the agent's ``digital body" to its ``social world": the other agents and services it must engage with to complete a complex task. For example, a coordinator agent may assign log‑collection to a context agent, ask a policy agent to verify constraints, and direct a tooling agent to execute approved steps, then integrate their outputs into a single recommendation. The Connection layer therefore concerns how agents communicate and coordinate, how they assign roles, delegate subtasks, and make trust decisions about whom to believe, when to verify, and when to escalate. Like the social layer of human society, it encodes identity, relationships, and the norms that enable cooperation, and it is also the place where manipulation can occur. %In the human analogy, Connection is the agent's ``social layer": identity, relationships, and norms that make cooperation possible, and also make manipulation possible. 
Communication is also the channel through which information and authority flow  through the system. So even if every agent has strong Core security, the overall system can still fail if messages are misleading, misattributed, or blindly trusted, because one compromised interaction can propagate through delegation chains \cite{hammond2025, pan2025why}.
%In the Core layer, we discussed an agent's dependence on system-level components (e.g., models, memory, tools, and execution environments). In the Connection layer, we shift attention to an agent's dependence on its ``social circle'', other agents and external services, with which it must interact, coordinate, and form judgments about trustworthiness. Multi-agent systems extend single-agent capability by enabling specialized agents to coordinate on complex objectives that are difficult for any single agent to execute reliably end-to-end. For example, a coordinator agent may delegate information gathering to a retrieval agent, request checks from a compliance agent, and assign execution steps to a tooling agent, before integrating intermediate output into an evidence-backed recommendation or action plan \cite{anthropic2025multiagent}. The Connection layer, therefore, captures the communicative and social fabric of agentic AI: how agents exchange messages, negotiate roles, and establish (or lose) trust while pursuing goals. Because communication is also the channel through which information and authority propagate, integrity and security of these interactions are central to safe operation, especially when an error, manipulation, or compromise in one agent can cascade through delegation chains.

\subheading{Threat Categories and Examples.} At the Connection layer, the reliability of any  outcome depends on the integrity of conversations, delegations, and agreements between agents and external endpoints they rely on \cite{He2025}. As such, trust becomes an operational necessity rather than a social courtesy. Each agent must judge whom to rely on, when to seek verification, and how much weight to grant the information that arrives through its channels of communication. Several threats exploit this interaction surface: (1) \textit{Rogue or malicious agents} can enter the ecosystem and inject misleading information while appearing cooperative \cite{motwani2025}. In the incident-response example, a malicious ``helper" agent could feed the coordinator a plausible but wrong root cause, steering remediation toward the wrong host. (2) \textit{Misinformation loops} can emerge when agents repeatedly consume and re-emit each other's outputs, creating cascading hallucinations where small errors become shared ``facts" \cite{Sapkota2025}. For incident response, one agent's incorrect summary of a log pattern can get repeated across agents until it looks like consensus. (3) \textit{Coordination and identity attacks} degrade reliability. Coordination manipulation can disrupt handoffs (for example, dropping a task, reordering steps, or redirecting a task to the wrong agent), while identity spoofing (impersonation) enables a counterfeit agent to pose as a trusted peer to gain access to channels, tools, or decision authority. In incident response, spoofing the oversight agent is especially dangerous because approvals are the last gate before high-impact actions. (4) \textit{Social engineering attacks} exploit norms of cooperation and delegation, such as ``be helpful" or ``follow the coordinator", to steer the group toward adversarial objectives. Together, these threats show how Connection-layer failures can undermine otherwise sound Core components by corrupting who agents trust, what they accept as evidence, and how tasks are coordinated.

\subheading{Key Remarks and Mitigation Implications.} The Connection layer shapes the pathways of communication: who may speak to whom, how tasks are woven together, and how agents come to understand one another’s goals and state \cite{qian2025}. Security at this layer must be designed and enforced as a whole, because it does not emerge simply by securing each agent on its own. Its central aim is governed trust: defining who may delegate, approve, or override actions; ensuring that messages can be traced to their source and audited; and preventing small missteps from cascading into failures at the group level. In practice, this requires strong authentication and authorization, delegation tied to clear roles with separation of duties, and structured messages that convey provenance—where a claim originated—along with supporting evidence and an explicit measure of confidence in its accuracy.  %It also requires mechanisms for verification and for resolving disagreements, so that repetition is never mistaken for correctness. 
It also requires verification and disagreement-resolution, so repetition is not mistaken for correctness.
When several agents echo the same claim without independent verification, the system should treat that pattern as a warning signal rather than as proof. Recent interoperability mechanisms such as Agent2Agent (A2A), Model Context Protocol (MCP), and function calling \cite{kong2025} can support these controls by making interactions more structured, attributable, and auditable.

\subsection{Cognition: Belief and Goal Integrity}

\subheading{Definition and Scope.} The Cognition layer captures the internal mechanisms that constitute an agent's ``digital mind": the processes that turn observations into beliefs, beliefs into goals, and goals into plans. It includes the agent's reasoning process, its internal representation of how the world works (world model), its belief state, reward and feedback signals, and planning system \cite{raza2025}. Together, these components determine how an agent interprets inputs, updates its understanding, prioritizes objectives, and decides what to do next. This layer is separated because cognition failures are not just communication or execution errors. They arise from how an agent forms, updates, and uses beliefs and objectives internally. Even with correct inputs and well-governed interactions, an agent can behave badly if its beliefs drift, its goals become misaligned, or its incentives favor ``looking successful" over being correct. Cognition therefore governs not only what the agent knows, but also what it is trying to optimize and how strongly it commits to its plan. As such, Cognition is referred as the agent's ``digital mind": attention, judgment, memory of what matters, and the ability to form a working belief about the current situation. Cognition is about whether the agent can think safely. In our incident-response running example, Cognition is what determines whether the triage agent interprets an alert as credible, whether the context agent forms the right hypothesis from logs, and whether the remediation agent selects a safe runbook step rather than an overreaction. If an agent's internal beliefs are wrong, it can take the wrong action with confidence, even when tools and communication channels are functioning as designed.

\subheading{Threat Categories and Examples.} Cognition-layer threats are mainly \textit{epistemic and motivational}: they distort what an agent believes, what it optimizes, and how it plans. Recent safety evaluations highlight cognition as a major risk source in goal-directed systems \cite{openai2024gpt4technicalreport, anthropic2025claude4systemcard}. We highlight five representative threats. (1) \textit{Belief drift} occurs when an agent's internal model diverges from reality due to corrupted inputs, biased feedback, or memory errors \cite{Huang2025}. If logs mislabel real intrusions as ``false alarms", a triage agent may start suppressing genuine alerts; in multi-agent settings, drift can propagate as agents reinforce one another's faulty inferences \cite{reid2025governedMultiAgentRiskAnalysis}. (2) \textit{Delusional reasoning} (agentic hallucination) is an internally coherent but false inference chain that becomes dangerous when stored in memory and reused as ``fact" \cite{Huang2025}. In incident response, an agent may assert a root cause without evidence and then treat it as confirmed context. (3) \textit{Reward hacking} arises when an agent optimizes proxy signals (e.g., ``fast resolution" or ``positive feedback") rather than the user's true objective \cite{Ji2025}. In operations, ``closing the ticket" can become the target instead of resolving the incident. (4) \textit{Unbounded autonomy} results when long-horizon planning lacks effective constraints, enabling costly or unsafe actions \cite{Gabriel2024}. A widely cited example is the July 2025 Replit coding-agent incident, where reports described unauthorized deletion of a production database during a ``code freeze" \cite{aiid1152}. (5) \textit{Internal goal hijacking} shifts the objective through targeted prompts or subtle memory edits while the agent still appears helpful \cite{Langosco2022}. For instance, a compromised remediation agent may prioritize ``restore service at any cost" over policies requiring staged containment and preservation of forensic evidence.

 \subheading{Key Remarks and Mitigation Implications.} Security at the Cognition layer concerns \textit{belief and goal integrity}: ensuring that an agent's internal representations remain tied to reality and its objectives remain aligned with user intent, even under uncertainty or operational stress. Cognition security must shape how the agent reasons, tracks its progress, and commits to plans, requiring safeguards embedded directly into its decision loop. In practice, this entails: (i) grounding and consistency checks that periodically re‑validate key beliefs against trusted sources to counter belief drift and delusional reasoning; (ii) improved success signals, combined with adversarial testing, to reduce reward hacking; (iii) bounded autonomy through budgets, forbidden‑action lists, escalation gates, and stop rules to prevent unsafe instrumental escalation; and (iv) protection of goals and memory via separated read/write permissions, signed or versioned policy states, and audits of changes that could otherwise shift what the agent optimizes. To reduce Cognition‑layer threats, policy‑guided methods such as Constitutional AI \cite{anthropic2023specific_general_constitutional_ai} constrain an agent's reasoning; supervision and self‑checks detect early signs of drift \cite{openai2025cotmonitorability}; and automated evaluations, along with continuous red‑teaming of long‑horizon behavior (e.g., OpenAI Evals \cite{openai-evals}), stress‑test agents under adversarial conditions.

\subsection{Compliance: Governance and Ethical Security}

\subheading{Definition and Scope.} The Compliance layer positions a multi‑agent system within the external norms and institutional controls that govern its behavior, including legal and regulatory requirements, organizational policies, audit and logging obligations, and ethical constraints on permissible goals, tools, and interaction patterns. Just as stable societies rely on rule‑based accountability and transparency to align individual actions with shared values, this layer provides the structural boundaries that ensure agents operate within acceptable limits \cite{Keping2018}, agentic AI systems require explicit governance structures to keep autonomous behavior within legal and ethical limits. The Compliance layer defines what the system \textit{is permitted} to do, under which conditions, with whose authorization, and with what evidentiary basis, ensuring that actions remain attributable, reviewable, and auditable. In practice, Compliance operationalizes governance through enforceable constraints: clearly specified allowed and prohibited actions, approval and escalation gates for sensitive operations, segregation of duties for high‑impact decisions, and requirements for transparency, retention, and post‑hoc review, all grounded in legal and normative regimes such as GDPR \cite{GDPR}, the EU AI Act \cite{eu2024aiact}, and the OECD AI Principles \cite{oecd2024g7toolkit}. These controls come under particular strain in agentic systems because they can construct multi‑step plans, invoke external tools, and coordinate with other agents over time, producing extended action chains with cumulative regulatory impact. In the incident‑response example, Compliance dictates which containment steps require human approval, which data sources agents may access, what must be logged, and who is accountable for the final decision. If permissions, logging, or escalation rules are misconfigured, the system may enact a sequence of actions with regulatory implications and limited auditability, even when each individual step appears locally justified.

\subheading{Threat Categories and Examples.}
Compliance-layer threats are governance failures: they occur when autonomy is not bounded by enforceable policy, incentives pull behavior away from norms, or oversight cannot detect and correct drift. These risks are amplified in agentic systems because agents can execute multi-step actions over time across many tools and services \cite{amodei2016,Gabriel2024}. We highlight four representative threat classes. (1) \textit{Misaligned autonomy} occurs when an agent acts outside its authorized scope because permissions, time limits, or approval gates are missing or not enforced. For example, in July~2025 Google's Gemini CLI coding assistant was reported to cause severe file loss after misinterpreting a failed directory-creation step and proceeding with file removal \cite{aiid1178}. (2) \textit{Ethical drift / unbounded optimization} arises when agents optimize local metrics (e.g., speed or ``get it done") over institutional intent; Project Vend illustrates how social pressure and open-ended objectives can drive loss-making actions such as giving away inventory \cite{anthropic2025}. (3) \textit{Guardrail evasion} occurs when agents bypass safety or policy checks through the toolchain. EchoLeak (CVE-2025-32711) shows a ``zero-click" prompt-injection chain against Microsoft~365 Copilot that enabled data exfiltration across trust boundaries \cite{reddy2025echoleakrealworldzeroclickprompt}. This is a Compliance failure because an organization may be policy-compliant on paper yet still leak data if guardrails are not enforceable across retrieval and tool execution. (4) \textit{Inadequate oversight and interfaces} arise when humans cannot reconstruct what happened (who authorized actions, which policies applied, and which tool calls executed), undermining audit, rollback, and accountability after high-impact behavior.

\subheading{Key Remarks and Mitigation Implications.}
Compliance‑layer security concerns \textit{enforceable guardrails and accountability}: converting policy into operational constraints, ensuring that actions are attributable and auditable—what occurred, under whose authority, and for what purpose—and sustaining lifecycle governance for continual assessment and revision. Its role is not to regulate capability or reasoning, but to govern legitimacy, responsibility, and demonstrable control. In practice, organizations increasingly anchor this layer in formal governance frameworks, such as ISO/IEC~42001 \cite{isoiec42001_2023}
 and the NIST AI Risk Management Framework \cite{nist2023airmf} and
 align deployments with applicable legal requirements (e.g., transparency and disclosure duties under the EU AI Act \cite{euaiact_article50}). or agentic systems, the prevailing model is \textit{bounded autonomy with oversight}: role‑based permissions and separation of duties for high‑impact operations, explicit approval and escalation gates, default logging of prompts, tool calls, and data access, third‑party and vendor due‑diligence requirements, and incident‑readiness measures for monitoring, escalation, and reporting. Together, these controls maintain organizational accountability as autonomy increases.

\section{Broader Implications for AI Security Research}
The emergence of agentic and multi‑agent AI systems challenges long‑standing assumptions in security research. As these systems reason, plan, coordinate, and act autonomously over extended horizons, security failures arise not only from isolated exploits but also from interactions among beliefs, goals, coordination mechanisms, and governance structures. The 4C framework offers a unified lens for analyzing these risks and organizing mitigations, clarifying how failures can originate at distinct layers and compound across them. In doing so, it shifts the focus from ``securing a model" to securing an evolving socio‑technical system.

%The rise of agentic and multi-agent AI systems challenges long-standing assumptions in security research. As systems gain the ability to reason, plan, coordinate, and act autonomously over extended horizons, security failures increasingly emerge not from isolated exploits, but from the interaction between beliefs, goals, coordination, and governance. The 4C framework—\emph{Core, Connection, Cognition, and Compliance}—offers a unifying lens for understanding these risks and for organizing both adversarial taxonomies and mitigation strategies. Rather than treating safety, alignment, and governance as adjacent concerns, the framework highlights how security failures can originate at different layers and compound across them.

%The rise of agentic and multi-agent AI systems challenges long-standing assumptions in security research. As these systems reason, plan, coordinate, and act autonomously on extended horizons, security failures increasingly stem not from isolated exploits but from interactions between beliefs, goals, coordination, and governance. The 4C framework—Core, Connection, Cognition and Compliance—provides a unifying lens for analyzing these risks and organizing mitigation strategies, highlighting how failures can originate at different layers and compound across them.

\subheading{From Asset Protection to Behavioral Integrity.} Traditional security focuses on protecting assets such as data, compute, interfaces, and infrastructure. While these remain essential at the Core and Connection layers, agentic AI introduces failure modes that can arise even when assets are secure and access is formally ``correct". An agent may use only authorized tools yet still cause harm if its beliefs drift, its goals misgeneralize, or its planning adopts undesirable instrumental strategies. These are failures of behavioral integrity, not perimeter defense. The 4C framework reflects this shift: Core ensures correct execution, Connection governs delegation and influence, Cognition preserves belief and goal integrity, and Compliance enforces norms and accountability. Security research must therefore go beyond robustness to ensure behavior remains aligned with intent over time, and develop adversarial taxonomies that include cognitive manipulation, coordination misuse, and governance breakdowns alongside traditional exploits.%Traditional security has centered on protecting assets such as data, compute, interfaces, and infrastructure. While these remain essential at the Core and Connection layers, agentic AI introduces broader failure modes that can arise even when assets are secure and access is formally ``correct". An agent may use only authorized tools and actions yet still produce harmful behavior if its beliefs drift, its goals misgeneralize, or its planning adopts undesirable instrumental strategies. These are failures of behavioral integrity rather than perimeter defense. The 4C framework reframes security around this point: Core assures correct execution, Connection governs delegation and influence, Cognition maintains belief and goal integrity, and Compliance enforces institutional norms and accountability. Security research must therefore move beyond robustness alone to ensuring that behavior across these layers remains aligned with intent over time. This shift also calls for richer adversarial taxonomies that encompass cognitive manipulation, coordination misuse, and governance breakdowns alongside traditional exploits.

\subheading{From Defense in Depth to Defense of Intentions.} Classic defense-in-depth assumes attackers penetrate from the outside. In agentic AI, many risks arise internally from the system's own reasoning and optimization. An agent may follow every local rule yet still violate the designer’s global intent, shifting the focus from defending systems to defending intent. In the 4C framework, intent must hold across Core (execution and access), Connection (influence, communication, delegation), Cognition (beliefs and objectives), and Compliance (what is permitted and auditable). Mitigation therefore need to be layered both technically and semantically, combining grounded beliefs, bounded planning, governed delegation, and policy-backed accountability. This also calls for benchmarks that stress-test long-horizon behavior, goal stability, and cross-layer failure modes.

\subheading{Humans as Agents in Mixed Socio-Technical Systems.} %An additional implication for AI security arises when humans function as agents within the same socio-technical ecosystem. In many deployments, humans delegate tasks, approve actions, provide feedback, and influence agent behavior, effectively participating in shared decision loops. Mixed human–agent systems introduce distinct risks: AI agents can amplify human vulnerabilities such as trust miscalibration or social engineering, while human actions—fatigue, over-reliance, or policy bypass—can propagate failures across agent networks. As a result, risk emerges not only from agent behavior, but from human–agent interaction dynamics. This blurs traditional boundaries between insider threats, social engineering, and agent misalignment, reinforcing the need for security models that account for influence, authority, and accountability across both human and artificial actors.
An additional implication arises when humans function as agents within the same socio-technical ecosystem. In many deployments, humans delegate tasks, approve actions, provide feedback, and influence agent behavior, effectively participating in shared decision loops. Mixed human–agent systems introduce distinct risks: AI agents can amplify human vulnerabilities such as trust miscalibration or social engineering, while human actions such as fatigue, over-reliance, or policy bypass, can propagate failures across the agents. Risk thus emerges not only from agent behavior, but from human–agent interaction dynamics, blurring boundaries between insider threats, social engineering, and agent misalignment.

\subheading{Toward an Interdisciplinary, Layered Security Agenda.}
%The challenges highlighted by the 4C framework cannot be addressed within a single discipline. Core defenses draw on cybersecurity and distributed systems; Connection draws on human-centric and social trust; Cognition-layer robustness depends on advances in machine learning, interpretability, and alignment; Compliance-layer safeguards require insights from governance, law, and organizational practice. Moreover, many emerging threats—such as social engineering among agents or incentive-driven ethical drift—sit at the intersection of technical systems and social dynamics.
%As a result, AI security research must become explicitly interdisciplinary and layered. The 4C framework provides a common structure for integrating insights across fields, enabling adversarial taxonomies that span execution, interaction, reasoning, and governance, and mitigation strategies that reinforce one another rather than operating in isolation. By using this framework as a guide, the research community can move toward security approaches that are not only technically robust, but also behaviorally aligned, socially informed, and institutionally grounded—qualities that will be essential as AI systems become more autonomous, persistent, and embedded in real-world decision-making.
The challenges surfaced by the 4C framework cannot be resolved within any single discipline. By adopting a society‑inspired model, the framework connects technical mechanisms with institutional controls. Core defenses draw on cybersecurity and distributed systems; Connection engages insights from human‑centric security and trust; Cognition relies on advances in machine learning, interpretability, and alignment; and Compliance draws on governance, law, and organizational practice. This paper offers a structural map of risks introduced by agentic AI rather than a simple catalog of vulnerabilities, emphasizing that many emerging threats lie at the intersection of technical and social dynamics. The 4C framework provides a common structure for integrating these perspectives, enabling security measures that reinforce one another across layers. This layered, interdisciplinary approach will be increasingly important as agentic AI becomes more autonomous and embedded in real-world decisions.

\section{Conclusion}
As AI systems move from being tools to operating as autonomous agents, security must be reimagined in ways that mirror how human societies govern power, trust, and accountability. This article argues that the most serious risks arise not only from technical exploits but from failures of behavior, coordination, intention, and governance. The 4C Framework—Core, Connection, Cognition, and Compliance—captures these dimensions by linking agentic risks to societal counterparts of capability, trust, belief, and oversight. By organizing adversarial taxonomies and mitigations across these layers, the framework clarifies how failures originate and propagate in agentic ecosystems and provides a principled basis for building AI systems that remain trustworthy and governable as autonomy increases. The framework is conceptual, and future work is required to operationalize it through metrics, benchmarks, and tools that can observe, audit, and constrain beliefs, goals, and long-horizon planning.

\bibliographystyle{ACM-Reference-Format}

%\bibliography{ref}
\end{document}